\documentclass[prl,showpacs,floatfix,twocolumn,superscriptaddress]{revtex4}
\usepackage{graphicx}
\usepackage{bm}
\usepackage{natbib}
\usepackage{float}
\usepackage{amsmath}
\usepackage{amssymb}
\usepackage{hyperref}
\usepackage{graphicx}

\newcommand{\etal}{\textit{et al.}}

\begin{document}
\preprint{BFS3}
\title{Magnetic-order-driven metal-insulator transitions in the \\
quasi-one-dimensional spin-ladder compounds BaFe$_2$S$_3$ and BaFe$_2$Se$_3$}

\author{Seulki Roh}
\thanks{These authors contributed equally.}
\affiliation{Department of Physics, Sungkyunkwan University, Suwon, Gyeonggi-do 16419, Republic of Korea}
\affiliation{1.Physikalisches Institut, Universit{\"a}t Stuttgart, 70569 Stuttgart, Germany}
\author{Soohyeon Shin}
\thanks{These authors contributed equally.}
\affiliation{Department of Physics, Sungkyunkwan University, Suwon, Gyeonggi-do 16419, Republic of Korea}
\affiliation{Center for Quantum Materials and Superconductivity (CQMS), Sungkyunkwan University, Suwon, Gyeonggi-do 16419, Republic of Korea}

\author{Jaekyung Jang}
\thanks{These authors contributed equally.}
\affiliation{Department of Physics, Sungkyunkwan University, Suwon, Gyeonggi-do 16419, Republic of Korea}
\author{Seokbae Lee}
\affiliation{Department of Physics, Sungkyunkwan University, Suwon, Gyeonggi-do 16419, Republic of Korea}
\author{Myounghoon Lee}
\affiliation{Department of Physics, Sungkyunkwan University, Suwon, Gyeonggi-do 16419, Republic of Korea}
\author{Yu-Seong Seo}
\affiliation{Department of Physics, Sungkyunkwan University, Suwon, Gyeonggi-do 16419, Republic of Korea}
\author{Weiwu Li}
\author{Tobias Biesner}
\author{Martin Dressel}
\affiliation{1.Physikalisches Institut, Universit{\"a}t Stuttgart, 70569 Stuttgart, Germany}
\author{Joo Yull Rhee}
\author{Tuson Park}
\affiliation{Department of Physics, Sungkyunkwan University, Suwon, Gyeonggi-do 16419, Republic of Korea}
\affiliation{Center for Quantum Materials and Superconductivity (CQMS), Sungkyunkwan University, Suwon, Gyeonggi-do 16419, Republic of Korea}
\author{Jungseek Hwang}
\email{jungseek@skku.edu}
\affiliation{Department of Physics, Sungkyunkwan University, Suwon, Gyeonggi-do 16419, Republic of Korea}

\date{\today}
\begin{abstract}
The quasi-one-dimensional spin ladder compounds, BaFe$_2$S$_3$ and BaFe$_2$Se$_3$, are investigated by infrared spectroscopy and density functional theory (DFT) calculations. We observe strong anisotropic electronic properties and an optical gap in the leg direction that is gradually filled above the antiferromagnetic (afm) ordering temperature, turning the systems into a metallic phase. Combining the optical data with the DFT calculations we associate the optical gap feature with the $p$-$d$ transition that appears only in the afm ordered state. Hence, the insulating ground state along the leg direction is attributed to Slater physics rather than Mott-type correlations.
\end{abstract}
\maketitle

Clarifying the nature of the electronic ground states of materials is a superior task in condensed matter physics as it is essential for understanding their physical properties. However, if more than one degrees of freedom (charge, spin, orbital and lattice) affect the electronic ground state and compete with each other, the situation becomes more complicated. In magnetic systems driving force of the insulating ground state can be either electronic correlations or magnetic order. The important debate is about which one plays the crucial role for stabilizing the insulating ground state observed at low temperatures. The detailed situation, of course, depends on the particular system and sometimes both cases are not exclusive \cite{Imada1998,Basov2011,Charnukha2017,Padilla2002,Perucchi2009,Vecchio2013,Sohn2015,Yamasaki2014,Watanabe2014}.

The quasi-one-dimensional spin ladder compounds BaFe$_2$S$_3$ (BFS) and BaFe$_2$Se$_3$ (BFSe) serve as models of this debate. The discovery of pressure-induced superconductivity in BFS opened a new category of the iron-based superconductors that does not contain the square lattice that was considered crucial for superconductivity in this family \cite{Hosono2018,Takahashi2015,Yamauchi2015}.  BFS has a quasi-one-dimensional antiferromagnetic (afm) spin-ladder structure along the Fe-ladder direction with a long-range afm ordering temperature $T_N \approx 120$ K. The Fe spins in the Fe-ladder direction of BFS are aligned in  alternating stacks of up-pairs and down-pairs of which spin ordering is called a CX-type antiferromagnet \cite{Popovic2015,Hirata2015,Takubo2017}. The BFSe shows a similar crystal structure; however, the Fe-ladder in BFSe  is slightly tilted compared to that in BFS \cite{Zhang2018,Svitlyk2019,Svitlyk2013,Ying2017}. Besides, the spin arrangement in the BFSe is also different from that in the BFS. The two pairs of up and down spins are alternating along the ladder of which spin ordering is called a block-type antiferromagnet. The reported $T_N$ varies from 140 to 250 K depending on the growth conditions
\cite{Mourigal2015,Nambu2012,Krzton-Maziopa2011,Lei2011,Saparov2011,Gao2017}.

Under pressure, both materials exhibit insulator-metal transitions (IMTs) that have been explained as a bandwidth-controlled Mott transition \cite{Takahashi2015,Yamauchi2015,Kobayashi2018,Mourigal2015,Rincon2014, Caron2012,Takubo2017,Luo2013,Zhang2017,Arita2015,Patel2016,Ying2017} since they are not accompanied by structural changes.
Moreover, recent photoemission spectroscopy \cite{Ootsuki2015} and resonant inelastic x-ray scattering \cite{Monney2013} studies reported coexistence of itinerant and localized electrons in BFSe, which was interpreted in terms of an orbital selective Mott phase \cite{Herbrych2018,Patel2019}.
Attempts to explain the orbital selective Mott phase by density matrix renormalization studies, however,
could not reproduce the insulating ground state of BFSe with the block-type afm ordering \cite{Herbrych2018}.
A more recent theoretical study with a slave spin technique revealed that the ground state of BFS below $T_N$ is not a Mott insulator but a strongly correlated Hund metal \cite{Pizarro2019}. The first-principles electronic-structure calculations with a generalized gradient approximation (GGA) on BFS \cite{Suzuki2015} and BFse \cite{Zhang2017,Zhang2018,Koley2018} using the exact spin configurations
reproduced the insulating ground state without including a Coulomb repulsion term $U$. These results infer that magnetic order could be decisive for the insulating ground states in BFS and BFSe. In other words, the Coulomb repulsion alone without consideration of the magnetic ordering is not sufficient to stabilize the insulating ground states observed in BFS and BFSe.

\begin{figure}
\includegraphics[width=\linewidth]{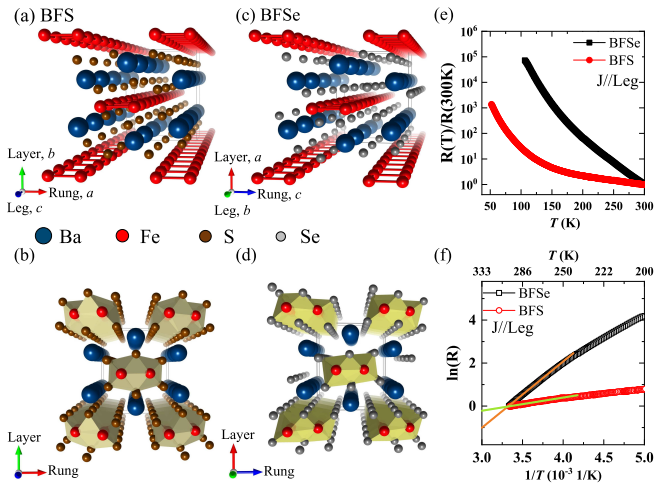}
\caption{(a-b) and 1(c-d) Schematic crystal structures of BaFe$_2$S$_3$ and BaFe$_2$Se$_3$ in different facets. (e) Measured dc resistivity data along the leg direction as functions of the temperature. (f) The Arrhenius plot, used to estimate the activation energy, $E_a$ with temperature range from 300 K to 260 K. The acquired $E_a$ of BFS and BFSe are 52.7 and 265 meV, respectively.}
\end{figure}

In the present study, we performed infrared spectroscopy and DFT calculations to elucidate the insulating electronic ground states of the title compounds. The calculations used the full potential and the exact spin configurations without including interactions. We observed insulator-metal transitions along the leg direction above the long-range afm ordering temperatures $T_N$, i.e.\ below the gap a Drude-type absorption builds up
by spectral weight transferred from higher energies. Our DFT calculations successfully reproduced the experimental results including the size of the optical gap, which only appears in the afm phase. Hence, we concluded that the insulating ground state of these two systems is driven by afm ordering, which could be understood by a scenario introduced by Slater \cite{Slater1951}. Our results are in stark contrast with those of previous studies and
bring new aspects to the understanding of the electronic ground states of these two interesting material systems.

High-quality single crystalline samples of BFS and BFSe were synthesized by a stoichiometric solid-state reaction method \cite{SM}. From magnetic susceptibility measurements we obtain $T_N= 100$ and 174~K for BFS and BFSe, respectively. To investigate the electronic structures, polarization-dependent optical reflectivity measurements were carried out in a wide energy range from 10 to 1500 meV, covering temperatures from 8 to 450~K.
From that the optical conductivity was calculated by the Kramers-Kronig relations \cite{Dressel2002,Tanner2019} and checked by variational dielectric function fitting \cite{Kuzmenko2019}.
For a better understanding of the measured spectra, we performed DFT calculations, using the WIEN2k package \cite{Blaha} implemented with a full-potential linearized-augmented-plane-wave method. To accommodate the correct afm configurations in the calculations, we used $2 \times 2\times 1$ and $1\times2\times2$ supercells for BFS and BFSe, respectively. The spin-orbit coupling was not included in the calculations. The lattice constants of BFS and BFSe were taken from well-documented data \cite{Hirata2015,Krzton-Maziopa2011}. As usual, we assumed only the direct transitions for the optical conductivity.

Fig. 1 depicts the crystal structures of BFS and BFSe. The different atomic sizes of S and Se result in distinct unit cell sizes and space groups: BFS belongs to the $Cmcm$ and BFSe to the $Pnma$ space groups. The iron ladders of BFSe along the layer direction are tilted compared with those of BFS \footnote{It is worthwhile to note that to avoid complications from different crystal axes between the two (BFS and BFSe) systems, we will denote the three axes as leg, layer, and rung ones. The corresponding crystal axes are, respectively, $c$-, $b$-, and $a$-axes for BFS and are, respectively, $b$-, $a$-, and $c$-axes for BFSe.}. Measured dc resistivity data of the two samples are displayed in Fig. 1(e) as a function of temperature. Both crystals manifest an insulating behavior from 300 K down to the lowest temperatures. In the Arrhenius plot of Fig. 1(f) the activation energies above $T$ = 260 K can be quickly estimated: $E_{a} =52.7$ and 265 meV for BFS and BFSe, respectively.

The reflectance spectra of BFS and BFSe along the leg directions are plotted in Fig. 2(a and b); for comparison we also display the 8-K spectra corresponding to the layer directions. While the latter polarization remains insulating at all temperatures \cite{SM},  along the legs the intensity in the low frequency region becomes enhanced as the temperature increases from 8 K. The corresponding conductivity spectra are displayed in Fig. 2(c) and 2(d). For BFS we observe a pronounced peak around 200 meV that keeps almost the same intensity up to $\approx 100$~K,
i.e.\ close  to $T_N$. When going 100 K  the peak diminishes and has completely disappeared at the highest temperature (450 K). In the case of BFSe, we observe basically the same temperature behavior albeit the peak shows up at $\sim$600 meV and the temperature scale is higher ($T_N =$ 174 K). The absorption edges $E_g$ of these peaks are comparable to the bandgaps estimated from activation energy ($E_g \approx 2E_a$) suggesting that these peaks correspond to excitations across the energy gap in the insulating state of the two samples. Note that at our highest temperature, 450 K, both samples manifest metallic characteristics with a finite conductivity at zero frequency captured by the Drude model.

\begin{figure}
\includegraphics[width=\linewidth]{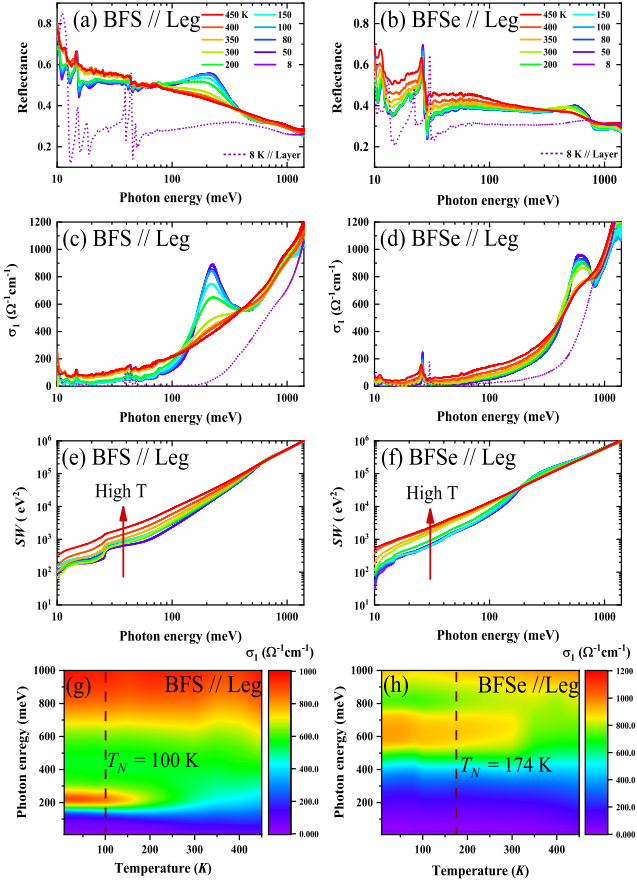}
	\centering
	\caption{Results of the optical spectroscopic study. (a) and (b) Measured reflectance spectrum of  BaFe$_2$S$_3$ and BaFe$_2$Se$_3$  along the leg direction, respectively. The purple dotted lines represent reflectance spectra along the layer direction at 8 K. (c) and (d) Optical conductivity ($\sigma_1$) of BFS and BFSe along the leg direction at various temperatures below and above $T_N$, respectively. (e) and (f) Corresponding spectral weight ($SW$) of BFS and BFSe, respectively. (d) and (h) Color-scaled maps of $\sigma_1$ of BFS and BFSe, respectively, as functions of temperature and photon energy. The vertical dashed lines indicate the long-range afm ordering temperatures, $T_N$.}
\end{figure}

Fig. 2(e and f) shows the (partial) spectral weight $SW(\omega) \equiv \int_0^\omega \sigma_1 (\omega^{\prime}) {\rm d}\omega^{\prime}$ of BFS and BFSe for different $T$. The overall spectral weight at various temperatures merge around 1 eV indicating that the number of electrons involved in the optical processes is conserved below 1 eV. At lower energies the $SW$ increases as the temperature rises, indicating a $SW$ transfer from high to low energies upon heating. In Fig. 2(g and h) we display color-scaled maps of $\sigma_1 (\omega)$ of both compounds along the leg direction as a function of temperature and photon energy. The gap-related peak of BFS near 200 meV is well-pronounced and saturated below $T_N$ marked as a vertical line; above $T_N$ it gradually disappears. The corresponding peak of BFSe exhibits a similar temperature-dependence with a higher $T_N$. The relation between the optical gap-related peak and $T_N$ demonstrates that the insulating ground states and the afm order along the leg direction are closely associated with each other.

The electronic density of states (DOS) and optical conductivity $\sigma _1$ obtained from the DFT calculations for various magnetic configurations are plotted in Fig. 3; panels (a and b) show the DOS of BFS and BFSe for ferromagnetic (fm) and non-magnetic (nm) cases. We always find a finite DOS at the Fermi energy ($E_F$), indicating that BFS and BFSe exhibit metallic ground states. The DOS for afm ordered phase -- the CX-type afm for BSE and the block-type afm for BFSe -- are displayed in Fig. 3(c and d): for both systems the afm order opens gaps at $E_F$ in the DOS resulting in insulating ground states. The calculated conductivity along the leg and layer directions is given in Fig. 3(e and f). The spectra manifest strong anisotropic electronic structures and fully agree with the measured $\sigma_1(\omega)$. Particularly, the peaks observed along the leg direction around 0.2 and 0.6 eV for BFS and BFSe, respectively, are well-reproduced by the DFT calculations. We identify these features as $p$-$d$ transitions \cite{SM}. Additionally, our results of BFS are in excellent agreement with reported results by GGA calculations \cite{Suzuki2015}.

\begin{figure}
\includegraphics[width=\linewidth]{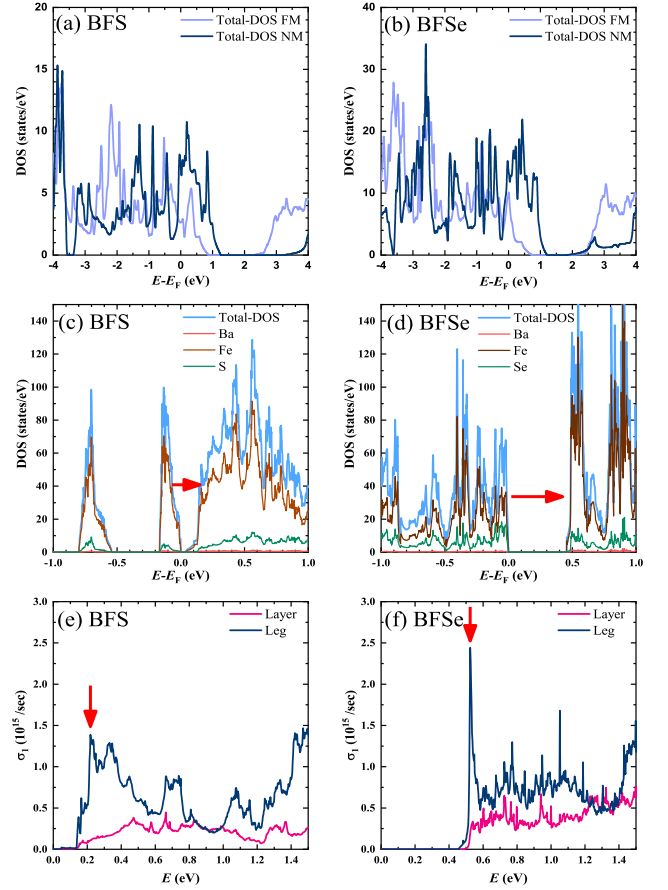}
	\centering
	\caption{(a) and (b) Density of states of BaFe$_2$S$_3$ and BaFe$_2$Se$_3$ for ferromagnetic (fm) and non-magnetic (nm) spin configurations. (c) and (d) Density of states of BFS and BFSe for their antiferromagnetic (afm) configurations: the CX-type afm for BSE and the block-type afm for BFSe. The red arrows represent the dominant interband transitions. (e) and (f) Calculated optical conductivity spectra of BFS and BFSe for their afm configurations along both the leg and layer directions. The red arrows mark the dominant absorption peaks, which are associated with the gap-related peaks.}
\end{figure}

Since the measured optical gaps are well-reproduced by the DFT calculations without including any electron-electron interaction $U$, we conclude that in these materials correlation effects play a minor role as a driving force for the insulating state. For both compounds, BFS and BFSe,  DFT calculations yield metallic ground states for fm and nm cases, while afm ordering always gives insulators. Obviously the electronic ground state of these two systems is more determined by magnetic configurations than by electronic correlations.
The fact that afm spin texture plays a crucial role for the insulating state is in full accord with Slater's scenario  \cite{Slater1951}. The proposed scenario is supported by the transfer of spectral weight into the sub-gap region above $T_N$ while the systems become insulating when magnetic order sets in.
Note that similar insulating states are found in other afm systems \cite{Kim2018,Vecchio2013,Charnukha2017,Padilla2002}.

Our explanation, however, is in contrast to the previously proposed Mott scenario \cite{Takahashi2015,Yamauchi2015,Kobayashi2018,Mourigal2015,Takubo2017,Li2016}
and the orbital selective Mott phase \cite{Herbrych2018,Patel2019}.
We believe that this discrepancy arises from the linear and peculiar spin texture of these compounds
that was not properly included in previous studies.
Our rigorous calculations with the full-potential and exact spin configurations unambiguously demonstrate
that the insulating ground states can be fully explained by Slater physics.
It is surprising that previous first-principles calculations never considered a Slater-type insulator
albeit the correct afm insulating ground state was reproduced without including any Hubbard-type potential $U$ \cite{Suzuki2015,Zhang2018,Koley2018,Saparov2011}.
When GGA calculation were combined with the dynamic-mean-field theory strong effects on correlation were proposed
\cite{Zhang2018}, although the Hubbard-type potentials $U = 0.5$~eV and $U^{\prime} = 0.1$~eV were rather small.

Usually, the Slater-type insulators undergo the IMT at $T_N$ \cite{Charnukha2017,Padilla2002,Perucchi2009,Vecchio2013}.
For BFS and BFSe, however, the IMT phase boundaries are not sharp and extended to temperatures above $T_N$;
also the metallic properties above $T_N$ are not well developed.
These features might be explained by the short-range afm order observed above $T_N$ \cite{Saparov2011,Popovic2015,Lei2011a,Caron2012,Mourigal2015,Nambu2012}.
Although long-range afm order is broken, the interplay between fluctuating short-range order and  quasi-one-dimensionality of the system may localize some of the electrons.
In both compounds itinerant and localized electrons coexist.
This picture is corroborated by the photoemission spectroscopic study of Ootsuki $\etal$ \cite{Ootsuki2015}
which was conducted at room temperature, i.e. well above $T_N$.
Our optical spectra indicate the coexistence of itinerant and localized electrons all the way down to $T_N$;
they appear as finite dc conductivity on the one hand and the gap-related peak on the other hand.
In contrast, the electronic state along the layer direction of both compounds stays insulating
at all measured temperatures; for more details see the  Supplemental Materials \cite{SM}.

Within the Mott picture the suppression of the insulating state by pressure was previously interpreted as bandwidth-tuned IMT \cite{Takahashi2015, Yamauchi2015}; it can be well explained using Slater physics.
Recent M{\"o}ssbauer spectroscopy under pressure, for instance, reported a sudden vanishing of the afm order in BFS at 9.9 GPa, a pressure similar to the value where IMT and superconductivity ($T_c$ = 24 K) simultaneously occur \cite{Materne2019,Takahashi2015,Yamauchi2015}. 
According to the Slater scenario the disappearance of afm order coincides with the IMT. 
Unlike the BFS, the sister compound BFSe undergoes several phase transitions under pressure 
including structural and magnetic ones \cite{Ying2017,Zhang2018,Svitlyk2019,Svitlyk2013}. 
Particular interesting transitions take place at 12 GPa, where the superconductivity ($T_c$ = 11 K) sets in 
when the afm transition takes place from the block to the CX-type. 
Here the specific spin textures and spin fluctuations in the vicinity of the magnetic transition could play a crucial role in both the IMT and superconductivity. In order to understand the electronic ground states in these intriguing quasi-one-dimensional spin ladder systems, we propose to seriously consider the role of magnetic ordering.

Based on our infrared study and DFT calculations we conclude strongly anisotropic electronic properties of the two quasi-one-dimensional spin ladder systems, BaFe$_2$S$_3$ and BaFe$_2$Se$_3$. 
The temperature-dependent infrared spectra reveal the IMT near the long-range afm ordering temperature $T_N$. 
With rising temperature spectral weight is transferred from the gap-related peak to the low-energy region.
The intimate relation between the optical gap and the afm ordering suggests that the afm ordering plays a crucial role in stabilizing the insulating ground state. 
DFT calculations of BFS (with the CX-type afm) and BFSe (with the block-type afm) fully reproduce our measured optical spectra in the insulating ground states without including electronic correlations. 
In contrast, the results of DFT calculations for the fm of nm cases of both compounds yield metallic ground states. 
Our comprehensive investigations clearly show that the observed insulating ground states in these quasi-one-dimensional spin ladder compounds are driven by afm orders, i.e.\ at low temperatures they are Slater-type insulators.

\acknowledgments We thank Ece Uykur and Yohei Saito for their fruitful discussions. J. Jang acknowledges support from NRF-2016R1A6A3A11933107. T. P. acknowledges financial support from the National Research Foundation of Korea (No. 2012R1A3A2048816). J. H. acknowledges financial support from the National Research Foundation of Korea (NRFK Grant No. 2017R1A2B4007387).

\bibliographystyle{apsrev4-1}
\bibliography{Ref}

\begin{thebibliography}{49}%
\makeatletter
\providecommand \@ifxundefined [1]{%
 \@ifx{#1\undefined}
}%
\providecommand \@ifnum [1]{%
 \ifnum #1\expandafter \@firstoftwo
 \else \expandafter \@secondoftwo
 \fi
}%
\providecommand \@ifx [1]{%
 \ifx #1\expandafter \@firstoftwo
 \else \expandafter \@secondoftwo
 \fi
}%
\providecommand \natexlab [1]{#1}%
\providecommand \enquote  [1]{``#1''}%
\providecommand \bibnamefont  [1]{#1}%
\providecommand \bibfnamefont [1]{#1}%
\providecommand \citenamefont [1]{#1}%
\providecommand \href@noop [0]{\@secondoftwo}%
\providecommand \href [0]{\begingroup \@sanitize@url \@href}%
\providecommand \@href[1]{\@@startlink{#1}\@@href}%
\providecommand \@@href[1]{\endgroup#1\@@endlink}%
\providecommand \@sanitize@url [0]{\catcode `\\12\catcode `\$12\catcode
  `\&12\catcode `\#12\catcode `\^12\catcode `\_12\catcode `\%12\relax}%
\providecommand \@@startlink[1]{}%
\providecommand \@@endlink[0]{}%
\providecommand \url  [0]{\begingroup\@sanitize@url \@url }%
\providecommand \@url [1]{\endgroup\@href {#1}{\urlprefix }}%
\providecommand \urlprefix  [0]{URL }%
\providecommand \Eprint [0]{\href }%
\providecommand \doibase [0]{http://dx.doi.org/}%
\providecommand \selectlanguage [0]{\@gobble}%
\providecommand \bibinfo  [0]{\@secondoftwo}%
\providecommand \bibfield  [0]{\@secondoftwo}%
\providecommand \translation [1]{[#1]}%
\providecommand \BibitemOpen [0]{}%
\providecommand \bibitemStop [0]{}%
\providecommand \bibitemNoStop [0]{.\EOS\space}%
\providecommand \EOS [0]{\spacefactor3000\relax}%
\providecommand \BibitemShut  [1]{\csname bibitem#1\endcsname}%
\let\auto@bib@innerbib\@empty
\bibitem [{\citenamefont {Imada}\ \emph {et~al.}(1998)\citenamefont {Imada},
  \citenamefont {Fujimori},\ and\ \citenamefont {Tokura}}]{Imada1998}%
  \BibitemOpen
  \bibfield  {author} {\bibinfo {author} {\bibfnamefont {M.}~\bibnamefont
  {Imada}}, \bibinfo {author} {\bibfnamefont {A.}~\bibnamefont {Fujimori}}, \
  and\ \bibinfo {author} {\bibfnamefont {Y.}~\bibnamefont {Tokura}},\ }\href
  {https://link.aps.org/doi/10.1103/RevModPhys.70.1039} {\bibfield  {journal}
  {\bibinfo  {journal} {Rev. Mod. Phys.}\ }\textbf {\bibinfo {volume} {70}},\
  \bibinfo {pages} {1039} (\bibinfo {year} {1998})}\BibitemShut {NoStop}%
\bibitem [{\citenamefont {Basov}\ \emph {et~al.}(2011)\citenamefont {Basov},
  \citenamefont {Averitt}, \citenamefont {van~der Marel}, \citenamefont
  {Dressel},\ and\ \citenamefont {Haule}}]{Basov2011}%
  \BibitemOpen
  \bibfield  {author} {\bibinfo {author} {\bibfnamefont {D.~N.}\ \bibnamefont
  {Basov}}, \bibinfo {author} {\bibfnamefont {R.~D.}\ \bibnamefont {Averitt}},
  \bibinfo {author} {\bibfnamefont {D.}~\bibnamefont {van~der Marel}}, \bibinfo
  {author} {\bibfnamefont {M.}~\bibnamefont {Dressel}}, \ and\ \bibinfo
  {author} {\bibfnamefont {K.}~\bibnamefont {Haule}},\ }\href
  {https://link.aps.org/doi/10.1103/RevModPhys.83.471} {\bibfield  {journal}
  {\bibinfo  {journal} {Rev. Mod. Phys.}\ }\textbf {\bibinfo {volume} {83}},\
  \bibinfo {pages} {471} (\bibinfo {year} {2011})}\BibitemShut {NoStop}%
\bibitem [{\citenamefont {Charnukha}\ \emph {et~al.}(2017)\citenamefont
  {Charnukha}, \citenamefont {Yin}, \citenamefont {Song}, \citenamefont {Cao},
  \citenamefont {Dai}, \citenamefont {Haule}, \citenamefont {Kotliar},\ and\
  \citenamefont {Basov}}]{Charnukha2017}%
  \BibitemOpen
  \bibfield  {author} {\bibinfo {author} {\bibfnamefont {A.}~\bibnamefont
  {Charnukha}}, \bibinfo {author} {\bibfnamefont {Z.~P.}\ \bibnamefont {Yin}},
  \bibinfo {author} {\bibfnamefont {Y.}~\bibnamefont {Song}}, \bibinfo {author}
  {\bibfnamefont {C.~D.}\ \bibnamefont {Cao}}, \bibinfo {author} {\bibfnamefont
  {P.}~\bibnamefont {Dai}}, \bibinfo {author} {\bibfnamefont {K.}~\bibnamefont
  {Haule}}, \bibinfo {author} {\bibfnamefont {G.}~\bibnamefont {Kotliar}}, \
  and\ \bibinfo {author} {\bibfnamefont {D.~N.}\ \bibnamefont {Basov}},\ }\href
  {\doibase 10.1103/PhysRevB.96.195121} {\bibfield  {journal} {\bibinfo
  {journal} {Phys. Rev. B}\ }\textbf {\bibinfo {volume} {96}},\ \bibinfo
  {pages} {195121} (\bibinfo {year} {2017})}\BibitemShut {NoStop}%
\bibitem [{\citenamefont {Padilla}\ \emph {et~al.}(2002)\citenamefont
  {Padilla}, \citenamefont {Mandrus},\ and\ \citenamefont
  {Basov}}]{Padilla2002}%
  \BibitemOpen
  \bibfield  {author} {\bibinfo {author} {\bibfnamefont {W.~J.}\ \bibnamefont
  {Padilla}}, \bibinfo {author} {\bibfnamefont {D.}~\bibnamefont {Mandrus}}, \
  and\ \bibinfo {author} {\bibfnamefont {D.~N.}\ \bibnamefont {Basov}},\ }\href
  {\doibase 10.1103/PhysRevB.66.035120} {\bibfield  {journal} {\bibinfo
  {journal} {Phys. Rev. B}\ }\textbf {\bibinfo {volume} {66}},\ \bibinfo
  {pages} {035120} (\bibinfo {year} {2002})}\BibitemShut {NoStop}%
\bibitem [{\citenamefont {Perucchi}\ \emph {et~al.}(2009)\citenamefont
  {Perucchi}, \citenamefont {Baldassarre}, \citenamefont {Postorino},\ and\
  \citenamefont {Lupi}}]{Perucchi2009}%
  \BibitemOpen
  \bibfield  {author} {\bibinfo {author} {\bibfnamefont {A.}~\bibnamefont
  {Perucchi}}, \bibinfo {author} {\bibfnamefont {L.}~\bibnamefont
  {Baldassarre}}, \bibinfo {author} {\bibfnamefont {P.}~\bibnamefont
  {Postorino}}, \ and\ \bibinfo {author} {\bibfnamefont {S.}~\bibnamefont
  {Lupi}},\ }\href {http://dx.doi.org/10.1088/0953-8984/21/32/323202}
  {\bibfield  {journal} {\bibinfo  {journal} {J. Phys.: Condens. Matter}\
  }\textbf {\bibinfo {volume} {21}},\ \bibinfo {pages} {323202} (\bibinfo
  {year} {2009})}\BibitemShut {NoStop}%
\bibitem [{\citenamefont {Vecchio}\ \emph {et~al.}(2013)\citenamefont
  {Vecchio}, \citenamefont {Perucchi}, \citenamefont {Di~Pietro}, \citenamefont
  {Limaj}, \citenamefont {Schade}, \citenamefont {Sun}, \citenamefont {Arai},
  \citenamefont {Yamaura},\ and\ \citenamefont {Lupi}}]{Vecchio2013}%
  \BibitemOpen
  \bibfield  {author} {\bibinfo {author} {\bibfnamefont {I.~L.}\ \bibnamefont
  {Vecchio}}, \bibinfo {author} {\bibfnamefont {A.}~\bibnamefont {Perucchi}},
  \bibinfo {author} {\bibfnamefont {P.}~\bibnamefont {Di~Pietro}}, \bibinfo
  {author} {\bibfnamefont {O.}~\bibnamefont {Limaj}}, \bibinfo {author}
  {\bibfnamefont {U.}~\bibnamefont {Schade}}, \bibinfo {author} {\bibfnamefont
  {Y.}~\bibnamefont {Sun}}, \bibinfo {author} {\bibfnamefont {M.}~\bibnamefont
  {Arai}}, \bibinfo {author} {\bibfnamefont {K.}~\bibnamefont {Yamaura}}, \
  and\ \bibinfo {author} {\bibfnamefont {S.}~\bibnamefont {Lupi}},\ }\href
  {https://doi.org/10.1038/srep02990} {\bibfield  {journal} {\bibinfo
  {journal} {Sci. Rep.}\ }\textbf {\bibinfo {volume} {3}},\ \bibinfo {pages}
  {2990} (\bibinfo {year} {2013})}\BibitemShut {NoStop}%
\bibitem [{\citenamefont {Sohn}\ \emph {et~al.}(2015)\citenamefont {Sohn},
  \citenamefont {Jeong}, \citenamefont {Jin}, \citenamefont {Kim},
  \citenamefont {Sandilands}, \citenamefont {Park}, \citenamefont {Kim},
  \citenamefont {Moon}, \citenamefont {Cho}, \citenamefont {Yamaura},
  \citenamefont {Hiroi},\ and\ \citenamefont {Noh}}]{Sohn2015}%
  \BibitemOpen
  \bibfield  {author} {\bibinfo {author} {\bibfnamefont {C.~H.}\ \bibnamefont
  {Sohn}}, \bibinfo {author} {\bibfnamefont {H.}~\bibnamefont {Jeong}},
  \bibinfo {author} {\bibfnamefont {H.}~\bibnamefont {Jin}}, \bibinfo {author}
  {\bibfnamefont {S.}~\bibnamefont {Kim}}, \bibinfo {author} {\bibfnamefont
  {L.~J.}\ \bibnamefont {Sandilands}}, \bibinfo {author} {\bibfnamefont
  {H.~J.}\ \bibnamefont {Park}}, \bibinfo {author} {\bibfnamefont {K.~W.}\
  \bibnamefont {Kim}}, \bibinfo {author} {\bibfnamefont {S.~J.}\ \bibnamefont
  {Moon}}, \bibinfo {author} {\bibfnamefont {D.-Y.}\ \bibnamefont {Cho}},
  \bibinfo {author} {\bibfnamefont {J.}~\bibnamefont {Yamaura}}, \bibinfo
  {author} {\bibfnamefont {Z.}~\bibnamefont {Hiroi}}, \ and\ \bibinfo {author}
  {\bibfnamefont {T.~W.}\ \bibnamefont {Noh}},\ }\href
  {https://link.aps.org/doi/10.1103/PhysRevLett.115.266402} {\bibfield
  {journal} {\bibinfo  {journal} {Phys. Rev. Lett.}\ }\textbf {\bibinfo
  {volume} {115}},\ \bibinfo {pages} {266402} (\bibinfo {year}
  {2015})}\BibitemShut {NoStop}%
\bibitem [{\citenamefont {Yamasaki}\ \emph {et~al.}(2014)\citenamefont
  {Yamasaki}, \citenamefont {Tachibana}, \citenamefont {Fujiwara},
  \citenamefont {Higashiya}, \citenamefont {Irizawa}, \citenamefont {Kirilmaz},
  \citenamefont {Pfaff}, \citenamefont {Scheiderer}, \citenamefont {Gabel},
  \citenamefont {Sing}, \citenamefont {Muro}, \citenamefont {Yabashi},
  \citenamefont {Tamasaku}, \citenamefont {Sato}, \citenamefont {Namatame},
  \citenamefont {Taniguchi}, \citenamefont {Hloskovskyy}, \citenamefont
  {Yoshida}, \citenamefont {Okabe}, \citenamefont {Isobe}, \citenamefont
  {Akimitsu}, \citenamefont {Drube}, \citenamefont {Claessen}, \citenamefont
  {Ishikawa}, \citenamefont {Imada}, \citenamefont {Sekiyama},\ and\
  \citenamefont {Suga}}]{Yamasaki2014}%
  \BibitemOpen
  \bibfield  {author} {\bibinfo {author} {\bibfnamefont {A.}~\bibnamefont
  {Yamasaki}}, \bibinfo {author} {\bibfnamefont {S.}~\bibnamefont {Tachibana}},
  \bibinfo {author} {\bibfnamefont {H.}~\bibnamefont {Fujiwara}}, \bibinfo
  {author} {\bibfnamefont {A.}~\bibnamefont {Higashiya}}, \bibinfo {author}
  {\bibfnamefont {A.}~\bibnamefont {Irizawa}}, \bibinfo {author} {\bibfnamefont
  {O.}~\bibnamefont {Kirilmaz}}, \bibinfo {author} {\bibfnamefont
  {F.}~\bibnamefont {Pfaff}}, \bibinfo {author} {\bibfnamefont
  {P.}~\bibnamefont {Scheiderer}}, \bibinfo {author} {\bibfnamefont
  {J.}~\bibnamefont {Gabel}}, \bibinfo {author} {\bibfnamefont
  {M.}~\bibnamefont {Sing}}, \bibinfo {author} {\bibfnamefont {T.}~\bibnamefont
  {Muro}}, \bibinfo {author} {\bibfnamefont {M.}~\bibnamefont {Yabashi}},
  \bibinfo {author} {\bibfnamefont {K.}~\bibnamefont {Tamasaku}}, \bibinfo
  {author} {\bibfnamefont {H.}~\bibnamefont {Sato}}, \bibinfo {author}
  {\bibfnamefont {H.}~\bibnamefont {Namatame}}, \bibinfo {author}
  {\bibfnamefont {M.}~\bibnamefont {Taniguchi}}, \bibinfo {author}
  {\bibfnamefont {A.}~\bibnamefont {Hloskovskyy}}, \bibinfo {author}
  {\bibfnamefont {H.}~\bibnamefont {Yoshida}}, \bibinfo {author} {\bibfnamefont
  {H.}~\bibnamefont {Okabe}}, \bibinfo {author} {\bibfnamefont
  {M.}~\bibnamefont {Isobe}}, \bibinfo {author} {\bibfnamefont
  {J.}~\bibnamefont {Akimitsu}}, \bibinfo {author} {\bibfnamefont
  {W.}~\bibnamefont {Drube}}, \bibinfo {author} {\bibfnamefont
  {R.}~\bibnamefont {Claessen}}, \bibinfo {author} {\bibfnamefont
  {T.}~\bibnamefont {Ishikawa}}, \bibinfo {author} {\bibfnamefont
  {S.}~\bibnamefont {Imada}}, \bibinfo {author} {\bibfnamefont
  {A.}~\bibnamefont {Sekiyama}}, \ and\ \bibinfo {author} {\bibfnamefont
  {S.}~\bibnamefont {Suga}},\ }\href {\doibase 10.1103/PhysRevB.89.121111}
  {\bibfield  {journal} {\bibinfo  {journal} {Phys. Rev. B}\ }\textbf {\bibinfo
  {volume} {89}},\ \bibinfo {pages} {121111} (\bibinfo {year}
  {2014})}\BibitemShut {NoStop}%
\bibitem [{\citenamefont {Watanabe}\ \emph {et~al.}(2014)\citenamefont
  {Watanabe}, \citenamefont {Shirakawa},\ and\ \citenamefont
  {Yunoki}}]{Watanabe2014}%
  \BibitemOpen
  \bibfield  {author} {\bibinfo {author} {\bibfnamefont {H.}~\bibnamefont
  {Watanabe}}, \bibinfo {author} {\bibfnamefont {T.}~\bibnamefont {Shirakawa}},
  \ and\ \bibinfo {author} {\bibfnamefont {S.}~\bibnamefont {Yunoki}},\ }\href
  {\doibase 10.1103/physrevb.89.165115} {\bibfield  {journal} {\bibinfo
  {journal} {Phys. Rev. B}\ }\textbf {\bibinfo {volume} {89}},\ \bibinfo
  {pages} {165115} (\bibinfo {year} {2014})}\BibitemShut {NoStop}%
\bibitem [{\citenamefont {Hosono}\ \emph {et~al.}(2018)\citenamefont {Hosono},
  \citenamefont {Yamamoto}, \citenamefont {Hiramatsu},\ and\ \citenamefont
  {Ma}}]{Hosono2018}%
  \BibitemOpen
  \bibfield  {author} {\bibinfo {author} {\bibfnamefont {H.}~\bibnamefont
  {Hosono}}, \bibinfo {author} {\bibfnamefont {A.}~\bibnamefont {Yamamoto}},
  \bibinfo {author} {\bibfnamefont {H.}~\bibnamefont {Hiramatsu}}, \ and\
  \bibinfo {author} {\bibfnamefont {Y.}~\bibnamefont {Ma}},\ }\href
  {http://www.sciencedirect.com/science/article/pii/S1369702117306545}
  {\bibfield  {journal} {\bibinfo  {journal} {Mater. Today}\ }\textbf {\bibinfo
  {volume} {21}},\ \bibinfo {pages} {278} (\bibinfo {year} {2018})}\BibitemShut
  {NoStop}%
\bibitem [{\citenamefont {Takahashi}\ \emph {et~al.}(2015)\citenamefont
  {Takahashi}, \citenamefont {Sugimoto}, \citenamefont {Nambu}, \citenamefont
  {Yamauchi}, \citenamefont {Hirata}, \citenamefont {Kawakami}, \citenamefont
  {Avdeev}, \citenamefont {Matsubayashi}, \citenamefont {Du}, \citenamefont
  {Kawashima}, \citenamefont {Soeda}, \citenamefont {Nakano}, \citenamefont
  {Uwatoko}, \citenamefont {Ueda}, \citenamefont {Sato},\ and\ \citenamefont
  {Ohgushi}}]{Takahashi2015}%
  \BibitemOpen
  \bibfield  {author} {\bibinfo {author} {\bibfnamefont {H.}~\bibnamefont
  {Takahashi}}, \bibinfo {author} {\bibfnamefont {A.}~\bibnamefont {Sugimoto}},
  \bibinfo {author} {\bibfnamefont {Y.}~\bibnamefont {Nambu}}, \bibinfo
  {author} {\bibfnamefont {T.}~\bibnamefont {Yamauchi}}, \bibinfo {author}
  {\bibfnamefont {Y.}~\bibnamefont {Hirata}}, \bibinfo {author} {\bibfnamefont
  {T.}~\bibnamefont {Kawakami}}, \bibinfo {author} {\bibfnamefont
  {M.}~\bibnamefont {Avdeev}}, \bibinfo {author} {\bibfnamefont
  {K.}~\bibnamefont {Matsubayashi}}, \bibinfo {author} {\bibfnamefont
  {F.}~\bibnamefont {Du}}, \bibinfo {author} {\bibfnamefont {C.}~\bibnamefont
  {Kawashima}}, \bibinfo {author} {\bibfnamefont {H.}~\bibnamefont {Soeda}},
  \bibinfo {author} {\bibfnamefont {S.}~\bibnamefont {Nakano}}, \bibinfo
  {author} {\bibfnamefont {Y.}~\bibnamefont {Uwatoko}}, \bibinfo {author}
  {\bibfnamefont {Y.}~\bibnamefont {Ueda}}, \bibinfo {author} {\bibfnamefont
  {T.~J.}\ \bibnamefont {Sato}}, \ and\ \bibinfo {author} {\bibfnamefont
  {K.}~\bibnamefont {Ohgushi}},\ }\href {\doibase 10.1038/nmat4351} {\bibfield
  {journal} {\bibinfo  {journal} {Nat. Mater.}\ }\textbf {\bibinfo {volume}
  {14}},\ \bibinfo {pages} {1008} (\bibinfo {year} {2015})}\BibitemShut
  {NoStop}%
\bibitem [{\citenamefont {Yamauchi}\ \emph {et~al.}(2015)\citenamefont
  {Yamauchi}, \citenamefont {Hirata}, \citenamefont {Ueda},\ and\ \citenamefont
  {Ohgushi}}]{Yamauchi2015}%
  \BibitemOpen
  \bibfield  {author} {\bibinfo {author} {\bibfnamefont {T.}~\bibnamefont
  {Yamauchi}}, \bibinfo {author} {\bibfnamefont {Y.}~\bibnamefont {Hirata}},
  \bibinfo {author} {\bibfnamefont {Y.}~\bibnamefont {Ueda}}, \ and\ \bibinfo
  {author} {\bibfnamefont {K.}~\bibnamefont {Ohgushi}},\ }\href
  {https://link.aps.org/doi/10.1103/PhysRevLett.115.246402} {\bibfield
  {journal} {\bibinfo  {journal} {Phys. Rev. Lett.}\ }\textbf {\bibinfo
  {volume} {115}},\ \bibinfo {pages} {246402} (\bibinfo {year}
  {2015})}\BibitemShut {NoStop}%
\bibitem [{\citenamefont {Popovi\ifmmode~\acute{c}\else \'{c}\fi{}}\ \emph
  {et~al.}(2015)\citenamefont {Popovi\ifmmode~\acute{c}\else \'{c}\fi{}},
  \citenamefont {\ifmmode \check{S}\else \v{S}\fi{}\ifmmode \acute{c}\else
  \'{c}\fi{}epanovi\ifmmode~\acute{c}\else \'{c}\fi{}}, \citenamefont
  {Lazarevi\ifmmode~\acute{c}\else \'{c}\fi{}}, \citenamefont {Opa\ifmmode
  \check{c}\else \v{c}\fi{}i\ifmmode~\acute{c}\else \'{c}\fi{}}, \citenamefont
  {Radonji\ifmmode~\acute{c}\else \'{c}\fi{}}, \citenamefont
  {Tanaskovi\ifmmode~\acute{c}\else \'{c}\fi{}}, \citenamefont {Lei},\ and\
  \citenamefont {Petrovic}}]{Popovic2015}%
  \BibitemOpen
  \bibfield  {author} {\bibinfo {author} {\bibfnamefont {Z.~V.}\ \bibnamefont
  {Popovi\ifmmode~\acute{c}\else \'{c}\fi{}}}, \bibinfo {author} {\bibfnamefont
  {M.}~\bibnamefont {\ifmmode \check{S}\else \v{S}\fi{}\ifmmode \acute{c}\else
  \'{c}\fi{}epanovi\ifmmode~\acute{c}\else \'{c}\fi{}}}, \bibinfo {author}
  {\bibfnamefont {N.}~\bibnamefont {Lazarevi\ifmmode~\acute{c}\else
  \'{c}\fi{}}}, \bibinfo {author} {\bibfnamefont {M.}~\bibnamefont {Opa\ifmmode
  \check{c}\else \v{c}\fi{}i\ifmmode~\acute{c}\else \'{c}\fi{}}}, \bibinfo
  {author} {\bibfnamefont {M.~M.}\ \bibnamefont {Radonji\ifmmode~\acute{c}\else
  \'{c}\fi{}}}, \bibinfo {author} {\bibfnamefont {D.}~\bibnamefont
  {Tanaskovi\ifmmode~\acute{c}\else \'{c}\fi{}}}, \bibinfo {author}
  {\bibfnamefont {H.}~\bibnamefont {Lei}}, \ and\ \bibinfo {author}
  {\bibfnamefont {C.}~\bibnamefont {Petrovic}},\ }\href {\doibase
  10.1103/PhysRevB.91.064303} {\bibfield  {journal} {\bibinfo  {journal} {Phys.
  Rev. B}\ }\textbf {\bibinfo {volume} {91}},\ \bibinfo {pages} {064303}
  (\bibinfo {year} {2015})}\BibitemShut {NoStop}%
\bibitem [{\citenamefont {Hirata}\ \emph {et~al.}(2015)\citenamefont {Hirata},
  \citenamefont {Maki}, \citenamefont {Yamaura}, \citenamefont {Yamauchi},\
  and\ \citenamefont {Ohgushi}}]{Hirata2015}%
  \BibitemOpen
  \bibfield  {author} {\bibinfo {author} {\bibfnamefont {Y.}~\bibnamefont
  {Hirata}}, \bibinfo {author} {\bibfnamefont {S.}~\bibnamefont {Maki}},
  \bibinfo {author} {\bibfnamefont {J.}~\bibnamefont {Yamaura}}, \bibinfo
  {author} {\bibfnamefont {T.}~\bibnamefont {Yamauchi}}, \ and\ \bibinfo
  {author} {\bibfnamefont {K.}~\bibnamefont {Ohgushi}},\ }\href
  {https://link.aps.org/doi/10.1103/PhysRevB.92.205109} {\bibfield  {journal}
  {\bibinfo  {journal} {Phys. Rev. B}\ }\textbf {\bibinfo {volume} {92}},\
  \bibinfo {pages} {205109} (\bibinfo {year} {2015})}\BibitemShut {NoStop}%
\bibitem [{\citenamefont {Takubo}\ \emph {et~al.}(2017)\citenamefont {Takubo},
  \citenamefont {Yokoyama}, \citenamefont {Wadati}, \citenamefont {Iwasaki},
  \citenamefont {Mizokawa}, \citenamefont {Boyko}, \citenamefont {Sutarto},
  \citenamefont {He}, \citenamefont {Hashizume}, \citenamefont {Imaizumi},
  \citenamefont {Aoyama}, \citenamefont {Imai},\ and\ \citenamefont
  {Ohgushi}}]{Takubo2017}%
  \BibitemOpen
  \bibfield  {author} {\bibinfo {author} {\bibfnamefont {K.}~\bibnamefont
  {Takubo}}, \bibinfo {author} {\bibfnamefont {Y.}~\bibnamefont {Yokoyama}},
  \bibinfo {author} {\bibfnamefont {H.}~\bibnamefont {Wadati}}, \bibinfo
  {author} {\bibfnamefont {S.}~\bibnamefont {Iwasaki}}, \bibinfo {author}
  {\bibfnamefont {T.}~\bibnamefont {Mizokawa}}, \bibinfo {author}
  {\bibfnamefont {T.}~\bibnamefont {Boyko}}, \bibinfo {author} {\bibfnamefont
  {R.}~\bibnamefont {Sutarto}}, \bibinfo {author} {\bibfnamefont
  {F.}~\bibnamefont {He}}, \bibinfo {author} {\bibfnamefont {K.}~\bibnamefont
  {Hashizume}}, \bibinfo {author} {\bibfnamefont {S.}~\bibnamefont {Imaizumi}},
  \bibinfo {author} {\bibfnamefont {T.}~\bibnamefont {Aoyama}}, \bibinfo
  {author} {\bibfnamefont {Y.}~\bibnamefont {Imai}}, \ and\ \bibinfo {author}
  {\bibfnamefont {K.}~\bibnamefont {Ohgushi}},\ }\href
  {https://link.aps.org/doi/10.1103/PhysRevB.96.115157} {\bibfield  {journal}
  {\bibinfo  {journal} {Phys. Rev. B}\ }\textbf {\bibinfo {volume} {96}},\
  \bibinfo {pages} {115157} (\bibinfo {year} {2017})}\BibitemShut {NoStop}%
\bibitem [{\citenamefont {Zhang}\ \emph {et~al.}(2018)\citenamefont {Zhang},
  \citenamefont {Lin}, \citenamefont {Zhang}, \citenamefont {Dagotto},\ and\
  \citenamefont {Dong}}]{Zhang2018}%
  \BibitemOpen
  \bibfield  {author} {\bibinfo {author} {\bibfnamefont {Y.}~\bibnamefont
  {Zhang}}, \bibinfo {author} {\bibfnamefont {L.-F.}\ \bibnamefont {Lin}},
  \bibinfo {author} {\bibfnamefont {J.-J.}\ \bibnamefont {Zhang}}, \bibinfo
  {author} {\bibfnamefont {E.}~\bibnamefont {Dagotto}}, \ and\ \bibinfo
  {author} {\bibfnamefont {S.}~\bibnamefont {Dong}},\ }\href
  {https://link.aps.org/doi/10.1103/PhysRevB.97.045119} {\bibfield  {journal}
  {\bibinfo  {journal} {Phys. Rev. B}\ }\textbf {\bibinfo {volume} {97}},\
  \bibinfo {pages} {045119} (\bibinfo {year} {2018})}\BibitemShut {NoStop}%
\bibitem [{\citenamefont {Svitlyk}\ \emph {et~al.}(2019)\citenamefont
  {Svitlyk}, \citenamefont {Garbarino}, \citenamefont {Rosa}, \citenamefont
  {Pomjakushina}, \citenamefont {Krzton-Maziopa}, \citenamefont {Conder},
  \citenamefont {Nunez-Regueiro},\ and\ \citenamefont {Mezouar}}]{Svitlyk2019}%
  \BibitemOpen
  \bibfield  {author} {\bibinfo {author} {\bibfnamefont {V.}~\bibnamefont
  {Svitlyk}}, \bibinfo {author} {\bibfnamefont {G.}~\bibnamefont {Garbarino}},
  \bibinfo {author} {\bibfnamefont {A.~D.}\ \bibnamefont {Rosa}}, \bibinfo
  {author} {\bibfnamefont {E.}~\bibnamefont {Pomjakushina}}, \bibinfo {author}
  {\bibfnamefont {A.}~\bibnamefont {Krzton-Maziopa}}, \bibinfo {author}
  {\bibfnamefont {K.}~\bibnamefont {Conder}}, \bibinfo {author} {\bibfnamefont
  {M.}~\bibnamefont {Nunez-Regueiro}}, \ and\ \bibinfo {author} {\bibfnamefont
  {M.}~\bibnamefont {Mezouar}},\ }\href
  {http://dx.doi.org/10.1088/1361-648X/aaf777} {\bibfield  {journal} {\bibinfo
  {journal} {J. Phys.: Condens. Matter}\ }\textbf {\bibinfo {volume} {31}},\
  \bibinfo {pages} {085401} (\bibinfo {year} {2019})}\BibitemShut {NoStop}%
\bibitem [{\citenamefont {Svitlyk}\ \emph {et~al.}(2013)\citenamefont
  {Svitlyk}, \citenamefont {Chernyshov}, \citenamefont {Pomjakushina},
  \citenamefont {Krzton-Maziopa}, \citenamefont {Conder}, \citenamefont
  {Pomjakushin}, \citenamefont {Pöttgen},\ and\ \citenamefont
  {Dmitriev}}]{Svitlyk2013}%
  \BibitemOpen
  \bibfield  {author} {\bibinfo {author} {\bibfnamefont {V.}~\bibnamefont
  {Svitlyk}}, \bibinfo {author} {\bibfnamefont {D.}~\bibnamefont {Chernyshov}},
  \bibinfo {author} {\bibfnamefont {E.}~\bibnamefont {Pomjakushina}}, \bibinfo
  {author} {\bibfnamefont {A.}~\bibnamefont {Krzton-Maziopa}}, \bibinfo
  {author} {\bibfnamefont {K.}~\bibnamefont {Conder}}, \bibinfo {author}
  {\bibfnamefont {V.}~\bibnamefont {Pomjakushin}}, \bibinfo {author}
  {\bibfnamefont {R.}~\bibnamefont {Pöttgen}}, \ and\ \bibinfo {author}
  {\bibfnamefont {V.}~\bibnamefont {Dmitriev}},\ }\href
  {http://dx.doi.org/10.1088/0953-8984/25/31/315403} {\bibfield  {journal}
  {\bibinfo  {journal} {J. Phys.: Condens. Matter}\ }\textbf {\bibinfo {volume}
  {25}},\ \bibinfo {pages} {315403} (\bibinfo {year} {2013})}\BibitemShut
  {NoStop}%
\bibitem [{\citenamefont {Ying}\ \emph {et~al.}(2017)\citenamefont {Ying},
  \citenamefont {Lei}, \citenamefont {Petrovic}, \citenamefont {Xiao},\ and\
  \citenamefont {Struzhkin}}]{Ying2017}%
  \BibitemOpen
  \bibfield  {author} {\bibinfo {author} {\bibfnamefont {J.}~\bibnamefont
  {Ying}}, \bibinfo {author} {\bibfnamefont {H.}~\bibnamefont {Lei}}, \bibinfo
  {author} {\bibfnamefont {C.}~\bibnamefont {Petrovic}}, \bibinfo {author}
  {\bibfnamefont {Y.}~\bibnamefont {Xiao}}, \ and\ \bibinfo {author}
  {\bibfnamefont {V.~V.}\ \bibnamefont {Struzhkin}},\ }\href
  {https://link.aps.org/doi/10.1103/PhysRevB.95.241109} {\bibfield  {journal}
  {\bibinfo  {journal} {Phys. Rev. B}\ }\textbf {\bibinfo {volume} {95}},\
  \bibinfo {pages} {241109} (\bibinfo {year} {2017})}\BibitemShut {NoStop}%
\bibitem [{\citenamefont {Mourigal}\ \emph {et~al.}(2015)\citenamefont
  {Mourigal}, \citenamefont {Wu}, \citenamefont {Stone}, \citenamefont
  {Neilson}, \citenamefont {Caron}, \citenamefont {McQueen},\ and\
  \citenamefont {Broholm}}]{Mourigal2015}%
  \BibitemOpen
  \bibfield  {author} {\bibinfo {author} {\bibfnamefont {M.}~\bibnamefont
  {Mourigal}}, \bibinfo {author} {\bibfnamefont {S.}~\bibnamefont {Wu}},
  \bibinfo {author} {\bibfnamefont {M.~B.}\ \bibnamefont {Stone}}, \bibinfo
  {author} {\bibfnamefont {J.~R.}\ \bibnamefont {Neilson}}, \bibinfo {author}
  {\bibfnamefont {J.~M.}\ \bibnamefont {Caron}}, \bibinfo {author}
  {\bibfnamefont {T.~M.}\ \bibnamefont {McQueen}}, \ and\ \bibinfo {author}
  {\bibfnamefont {C.~L.}\ \bibnamefont {Broholm}},\ }\href
  {https://link.aps.org/doi/10.1103/PhysRevLett.115.047401} {\bibfield
  {journal} {\bibinfo  {journal} {Phys. Rev. Lett.}\ }\textbf {\bibinfo
  {volume} {115}},\ \bibinfo {pages} {047401} (\bibinfo {year}
  {2015})}\BibitemShut {NoStop}%
\bibitem [{\citenamefont {Nambu}\ \emph {et~al.}(2012)\citenamefont {Nambu},
  \citenamefont {Ohgushi}, \citenamefont {Suzuki}, \citenamefont {Du},
  \citenamefont {Avdeev}, \citenamefont {Uwatoko}, \citenamefont {Munakata},
  \citenamefont {Fukazawa}, \citenamefont {Chi}, \citenamefont {Ueda},\ and\
  \citenamefont {Sato}}]{Nambu2012}%
  \BibitemOpen
  \bibfield  {author} {\bibinfo {author} {\bibfnamefont {Y.}~\bibnamefont
  {Nambu}}, \bibinfo {author} {\bibfnamefont {K.}~\bibnamefont {Ohgushi}},
  \bibinfo {author} {\bibfnamefont {S.}~\bibnamefont {Suzuki}}, \bibinfo
  {author} {\bibfnamefont {F.}~\bibnamefont {Du}}, \bibinfo {author}
  {\bibfnamefont {M.}~\bibnamefont {Avdeev}}, \bibinfo {author} {\bibfnamefont
  {Y.}~\bibnamefont {Uwatoko}}, \bibinfo {author} {\bibfnamefont
  {K.}~\bibnamefont {Munakata}}, \bibinfo {author} {\bibfnamefont
  {H.}~\bibnamefont {Fukazawa}}, \bibinfo {author} {\bibfnamefont
  {S.}~\bibnamefont {Chi}}, \bibinfo {author} {\bibfnamefont {Y.}~\bibnamefont
  {Ueda}}, \ and\ \bibinfo {author} {\bibfnamefont {T.~J.}\ \bibnamefont
  {Sato}},\ }\href {https://link.aps.org/doi/10.1103/PhysRevB.85.064413}
  {\bibfield  {journal} {\bibinfo  {journal} {Phys. Rev. B}\ }\textbf {\bibinfo
  {volume} {85}},\ \bibinfo {pages} {064413} (\bibinfo {year}
  {2012})}\BibitemShut {NoStop}%
\bibitem [{\citenamefont {Krzton-Maziopa}\ \emph {et~al.}(2011)\citenamefont
  {Krzton-Maziopa}, \citenamefont {Pomjakushina}, \citenamefont {Pomjakushin},
  \citenamefont {Sheptyakov}, \citenamefont {Chernyshov}, \citenamefont
  {Svitlyk},\ and\ \citenamefont {Conder}}]{Krzton-Maziopa2011}%
  \BibitemOpen
  \bibfield  {author} {\bibinfo {author} {\bibfnamefont {A.}~\bibnamefont
  {Krzton-Maziopa}}, \bibinfo {author} {\bibfnamefont {E.}~\bibnamefont
  {Pomjakushina}}, \bibinfo {author} {\bibfnamefont {V.}~\bibnamefont
  {Pomjakushin}}, \bibinfo {author} {\bibfnamefont {D.}~\bibnamefont
  {Sheptyakov}}, \bibinfo {author} {\bibfnamefont {D.}~\bibnamefont
  {Chernyshov}}, \bibinfo {author} {\bibfnamefont {V.}~\bibnamefont {Svitlyk}},
  \ and\ \bibinfo {author} {\bibfnamefont {K.}~\bibnamefont {Conder}},\ }\href
  {http://dx.doi.org/10.1088/0953-8984/23/40/402201} {\bibfield  {journal}
  {\bibinfo  {journal} {J. Phys.: Condens. Matter}\ }\textbf {\bibinfo {volume}
  {23}},\ \bibinfo {pages} {402201} (\bibinfo {year} {2011})}\BibitemShut
  {NoStop}%
\bibitem [{\citenamefont {Lei}\ \emph {et~al.}(2011{\natexlab{a}})\citenamefont
  {Lei}, \citenamefont {Ryu}, \citenamefont {Frenkel},\ and\ \citenamefont
  {Petrovic}}]{Lei2011}%
  \BibitemOpen
  \bibfield  {author} {\bibinfo {author} {\bibfnamefont {H.}~\bibnamefont
  {Lei}}, \bibinfo {author} {\bibfnamefont {H.}~\bibnamefont {Ryu}}, \bibinfo
  {author} {\bibfnamefont {A.~I.}\ \bibnamefont {Frenkel}}, \ and\ \bibinfo
  {author} {\bibfnamefont {C.}~\bibnamefont {Petrovic}},\ }\href {\doibase
  10.1103/PhysRevB.84.214511} {\bibfield  {journal} {\bibinfo  {journal} {Phys.
  Rev. B}\ }\textbf {\bibinfo {volume} {84}},\ \bibinfo {pages} {214511}
  (\bibinfo {year} {2011}{\natexlab{a}})}\BibitemShut {NoStop}%
\bibitem [{\citenamefont {Saparov}\ \emph {et~al.}(2011)\citenamefont
  {Saparov}, \citenamefont {Calder}, \citenamefont {Sipos}, \citenamefont
  {Cao}, \citenamefont {Chi}, \citenamefont {Singh}, \citenamefont
  {Christianson}, \citenamefont {Lumsden},\ and\ \citenamefont
  {Sefat}}]{Saparov2011}%
  \BibitemOpen
  \bibfield  {author} {\bibinfo {author} {\bibfnamefont {B.}~\bibnamefont
  {Saparov}}, \bibinfo {author} {\bibfnamefont {S.}~\bibnamefont {Calder}},
  \bibinfo {author} {\bibfnamefont {B.}~\bibnamefont {Sipos}}, \bibinfo
  {author} {\bibfnamefont {H.}~\bibnamefont {Cao}}, \bibinfo {author}
  {\bibfnamefont {S.}~\bibnamefont {Chi}}, \bibinfo {author} {\bibfnamefont
  {D.~J.}\ \bibnamefont {Singh}}, \bibinfo {author} {\bibfnamefont {A.~D.}\
  \bibnamefont {Christianson}}, \bibinfo {author} {\bibfnamefont {M.~D.}\
  \bibnamefont {Lumsden}}, \ and\ \bibinfo {author} {\bibfnamefont {A.~S.}\
  \bibnamefont {Sefat}},\ }\href
  {https://link.aps.org/doi/10.1103/PhysRevB.84.245132} {\bibfield  {journal}
  {\bibinfo  {journal} {Phys. Rev. B}\ }\textbf {\bibinfo {volume} {84}},\
  \bibinfo {pages} {245132} (\bibinfo {year} {2011})}\BibitemShut {NoStop}%
\bibitem [{\citenamefont {Gao}\ \emph {et~al.}(2017)\citenamefont {Gao},
  \citenamefont {Teng}, \citenamefont {Liu}, \citenamefont {Chen},
  \citenamefont {Tong}, \citenamefont {Li}, \citenamefont {Zhao},\ and\
  \citenamefont {Liu}}]{Gao2017}%
  \BibitemOpen
  \bibfield  {author} {\bibinfo {author} {\bibfnamefont {J.}~\bibnamefont
  {Gao}}, \bibinfo {author} {\bibfnamefont {Y.}~\bibnamefont {Teng}}, \bibinfo
  {author} {\bibfnamefont {W.}~\bibnamefont {Liu}}, \bibinfo {author}
  {\bibfnamefont {S.}~\bibnamefont {Chen}}, \bibinfo {author} {\bibfnamefont
  {W.}~\bibnamefont {Tong}}, \bibinfo {author} {\bibfnamefont {M.}~\bibnamefont
  {Li}}, \bibinfo {author} {\bibfnamefont {X.}~\bibnamefont {Zhao}}, \ and\
  \bibinfo {author} {\bibfnamefont {X.}~\bibnamefont {Liu}},\ }\href
  {http://dx.doi.org/10.1039/C7RA03031B} {\bibfield  {journal} {\bibinfo
  {journal} {RSC Adv.}\ }\textbf {\bibinfo {volume} {7}},\ \bibinfo {pages}
  {30433} (\bibinfo {year} {2017})}\BibitemShut {NoStop}%
\bibitem [{\citenamefont {Kobayashi}\ \emph {et~al.}(2018)\citenamefont
  {Kobayashi}, \citenamefont {Maki}, \citenamefont {Murakami}, \citenamefont
  {Hirata}, \citenamefont {Ohgushi},\ and\ \citenamefont
  {Yamaura}}]{Kobayashi2018}%
  \BibitemOpen
  \bibfield  {author} {\bibinfo {author} {\bibfnamefont {K.}~\bibnamefont
  {Kobayashi}}, \bibinfo {author} {\bibfnamefont {S.}~\bibnamefont {Maki}},
  \bibinfo {author} {\bibfnamefont {Y.}~\bibnamefont {Murakami}}, \bibinfo
  {author} {\bibfnamefont {Y.}~\bibnamefont {Hirata}}, \bibinfo {author}
  {\bibfnamefont {K.}~\bibnamefont {Ohgushi}}, \ and\ \bibinfo {author}
  {\bibfnamefont {J.}~\bibnamefont {Yamaura}},\ }\href
  {http://dx.doi.org/10.1088/1361-6668/aad790} {\bibfield  {journal} {\bibinfo
  {journal} {Supercond. Sci. Technol.}\ }\textbf {\bibinfo {volume} {31}},\
  \bibinfo {pages} {105002} (\bibinfo {year} {2018})}\BibitemShut {NoStop}%
\bibitem [{\citenamefont {Rincón}\ \emph {et~al.}(2014)\citenamefont
  {Rincón}, \citenamefont {Moreo}, \citenamefont {Alvarez},\ and\
  \citenamefont {Dagotto}}]{Rincon2014}%
  \BibitemOpen
  \bibfield  {author} {\bibinfo {author} {\bibfnamefont {J.}~\bibnamefont
  {Rincón}}, \bibinfo {author} {\bibfnamefont {A.}~\bibnamefont {Moreo}},
  \bibinfo {author} {\bibfnamefont {G.}~\bibnamefont {Alvarez}}, \ and\
  \bibinfo {author} {\bibfnamefont {E.}~\bibnamefont {Dagotto}},\ }\href
  {https://link.aps.org/doi/10.1103/PhysRevLett.112.106405} {\bibfield
  {journal} {\bibinfo  {journal} {Phys. Rev. Lett.}\ }\textbf {\bibinfo
  {volume} {112}},\ \bibinfo {pages} {106405} (\bibinfo {year}
  {2014})}\BibitemShut {NoStop}%
\bibitem [{\citenamefont {Caron}\ \emph {et~al.}(2012)\citenamefont {Caron},
  \citenamefont {Neilson}, \citenamefont {Miller}, \citenamefont {Arpino},
  \citenamefont {Llobet},\ and\ \citenamefont {McQueen}}]{Caron2012}%
  \BibitemOpen
  \bibfield  {author} {\bibinfo {author} {\bibfnamefont {J.~M.}\ \bibnamefont
  {Caron}}, \bibinfo {author} {\bibfnamefont {J.~R.}\ \bibnamefont {Neilson}},
  \bibinfo {author} {\bibfnamefont {D.~C.}\ \bibnamefont {Miller}}, \bibinfo
  {author} {\bibfnamefont {K.}~\bibnamefont {Arpino}}, \bibinfo {author}
  {\bibfnamefont {A.}~\bibnamefont {Llobet}}, \ and\ \bibinfo {author}
  {\bibfnamefont {T.~M.}\ \bibnamefont {McQueen}},\ }\href
  {https://link.aps.org/doi/10.1103/PhysRevB.85.180405} {\bibfield  {journal}
  {\bibinfo  {journal} {Phys. Rev. B}\ }\textbf {\bibinfo {volume} {85}},\
  \bibinfo {pages} {180405} (\bibinfo {year} {2012})}\BibitemShut {NoStop}%
\bibitem [{\citenamefont {Luo}\ \emph {et~al.}(2013)\citenamefont {Luo},
  \citenamefont {Nicholson}, \citenamefont {Rincón}, \citenamefont {Liang},
  \citenamefont {Riera}, \citenamefont {Alvarez}, \citenamefont {Wang},
  \citenamefont {Ku}, \citenamefont {Samolyuk}, \citenamefont {Moreo},\ and\
  \citenamefont {Dagotto}}]{Luo2013}%
  \BibitemOpen
  \bibfield  {author} {\bibinfo {author} {\bibfnamefont {Q.}~\bibnamefont
  {Luo}}, \bibinfo {author} {\bibfnamefont {A.}~\bibnamefont {Nicholson}},
  \bibinfo {author} {\bibfnamefont {J.}~\bibnamefont {Rincón}}, \bibinfo
  {author} {\bibfnamefont {S.}~\bibnamefont {Liang}}, \bibinfo {author}
  {\bibfnamefont {J.}~\bibnamefont {Riera}}, \bibinfo {author} {\bibfnamefont
  {G.}~\bibnamefont {Alvarez}}, \bibinfo {author} {\bibfnamefont
  {L.}~\bibnamefont {Wang}}, \bibinfo {author} {\bibfnamefont {W.}~\bibnamefont
  {Ku}}, \bibinfo {author} {\bibfnamefont {G.~D.}\ \bibnamefont {Samolyuk}},
  \bibinfo {author} {\bibfnamefont {A.}~\bibnamefont {Moreo}}, \ and\ \bibinfo
  {author} {\bibfnamefont {E.}~\bibnamefont {Dagotto}},\ }\href
  {https://link.aps.org/doi/10.1103/PhysRevB.87.024404} {\bibfield  {journal}
  {\bibinfo  {journal} {Phys. Rev. B}\ }\textbf {\bibinfo {volume} {87}},\
  \bibinfo {pages} {024404} (\bibinfo {year} {2013})}\BibitemShut {NoStop}%
\bibitem [{\citenamefont {Zhang}\ \emph {et~al.}(2017)\citenamefont {Zhang},
  \citenamefont {Lin}, \citenamefont {Zhang}, \citenamefont {Dagotto},\ and\
  \citenamefont {Dong}}]{Zhang2017}%
  \BibitemOpen
  \bibfield  {author} {\bibinfo {author} {\bibfnamefont {Y.}~\bibnamefont
  {Zhang}}, \bibinfo {author} {\bibfnamefont {L.}~\bibnamefont {Lin}}, \bibinfo
  {author} {\bibfnamefont {J.-J.}\ \bibnamefont {Zhang}}, \bibinfo {author}
  {\bibfnamefont {E.}~\bibnamefont {Dagotto}}, \ and\ \bibinfo {author}
  {\bibfnamefont {S.}~\bibnamefont {Dong}},\ }\href
  {https://link.aps.org/doi/10.1103/PhysRevB.95.115154} {\bibfield  {journal}
  {\bibinfo  {journal} {Phys. Rev. B}\ }\textbf {\bibinfo {volume} {95}},\
  \bibinfo {pages} {115154} (\bibinfo {year} {2017})}\BibitemShut {NoStop}%
\bibitem [{\citenamefont {Arita}\ \emph {et~al.}(2015)\citenamefont {Arita},
  \citenamefont {Ikeda}, \citenamefont {Sakai},\ and\ \citenamefont
  {Suzuki}}]{Arita2015}%
  \BibitemOpen
  \bibfield  {author} {\bibinfo {author} {\bibfnamefont {R.}~\bibnamefont
  {Arita}}, \bibinfo {author} {\bibfnamefont {H.}~\bibnamefont {Ikeda}},
  \bibinfo {author} {\bibfnamefont {S.}~\bibnamefont {Sakai}}, \ and\ \bibinfo
  {author} {\bibfnamefont {M.-T.}\ \bibnamefont {Suzuki}},\ }\href
  {https://link.aps.org/doi/10.1103/PhysRevB.92.054515} {\bibfield  {journal}
  {\bibinfo  {journal} {Phys. Rev. B}\ }\textbf {\bibinfo {volume} {92}},\
  \bibinfo {pages} {054515} (\bibinfo {year} {2015})}\BibitemShut {NoStop}%
\bibitem [{\citenamefont {Patel}\ \emph {et~al.}(2016)\citenamefont {Patel},
  \citenamefont {Nocera}, \citenamefont {Alvarez}, \citenamefont {Arita},
  \citenamefont {Moreo},\ and\ \citenamefont {Dagotto}}]{Patel2016}%
  \BibitemOpen
  \bibfield  {author} {\bibinfo {author} {\bibfnamefont {N.~D.}\ \bibnamefont
  {Patel}}, \bibinfo {author} {\bibfnamefont {A.}~\bibnamefont {Nocera}},
  \bibinfo {author} {\bibfnamefont {G.}~\bibnamefont {Alvarez}}, \bibinfo
  {author} {\bibfnamefont {R.}~\bibnamefont {Arita}}, \bibinfo {author}
  {\bibfnamefont {A.}~\bibnamefont {Moreo}}, \ and\ \bibinfo {author}
  {\bibfnamefont {E.}~\bibnamefont {Dagotto}},\ }\href
  {https://link.aps.org/doi/10.1103/PhysRevB.94.075119} {\bibfield  {journal}
  {\bibinfo  {journal} {Phys. Rev. B}\ }\textbf {\bibinfo {volume} {94}},\
  \bibinfo {pages} {075119} (\bibinfo {year} {2016})}\BibitemShut {NoStop}%
\bibitem [{\citenamefont {Ootsuki}\ \emph {et~al.}(2015)\citenamefont
  {Ootsuki}, \citenamefont {Saini}, \citenamefont {Du}, \citenamefont {Hirata},
  \citenamefont {Ohgushi}, \citenamefont {Ueda},\ and\ \citenamefont
  {Mizokawa}}]{Ootsuki2015}%
  \BibitemOpen
  \bibfield  {author} {\bibinfo {author} {\bibfnamefont {D.}~\bibnamefont
  {Ootsuki}}, \bibinfo {author} {\bibfnamefont {N.~L.}\ \bibnamefont {Saini}},
  \bibinfo {author} {\bibfnamefont {F.}~\bibnamefont {Du}}, \bibinfo {author}
  {\bibfnamefont {Y.}~\bibnamefont {Hirata}}, \bibinfo {author} {\bibfnamefont
  {K.}~\bibnamefont {Ohgushi}}, \bibinfo {author} {\bibfnamefont
  {Y.}~\bibnamefont {Ueda}}, \ and\ \bibinfo {author} {\bibfnamefont
  {T.}~\bibnamefont {Mizokawa}},\ }\href
  {https://link.aps.org/doi/10.1103/PhysRevB.91.014505} {\bibfield  {journal}
  {\bibinfo  {journal} {Phys. Rev. B}\ }\textbf {\bibinfo {volume} {91}},\
  \bibinfo {pages} {014505} (\bibinfo {year} {2015})}\BibitemShut {NoStop}%
\bibitem [{\citenamefont {Monney}\ \emph {et~al.}(2013)\citenamefont {Monney},
  \citenamefont {Uldry}, \citenamefont {Zhou}, \citenamefont {Krzton-Maziopa},
  \citenamefont {Pomjakushina}, \citenamefont {Strocov}, \citenamefont
  {Delley},\ and\ \citenamefont {Schmitt}}]{Monney2013}%
  \BibitemOpen
  \bibfield  {author} {\bibinfo {author} {\bibfnamefont {C.}~\bibnamefont
  {Monney}}, \bibinfo {author} {\bibfnamefont {A.}~\bibnamefont {Uldry}},
  \bibinfo {author} {\bibfnamefont {K.~J.}\ \bibnamefont {Zhou}}, \bibinfo
  {author} {\bibfnamefont {A.}~\bibnamefont {Krzton-Maziopa}}, \bibinfo
  {author} {\bibfnamefont {E.}~\bibnamefont {Pomjakushina}}, \bibinfo {author}
  {\bibfnamefont {V.~N.}\ \bibnamefont {Strocov}}, \bibinfo {author}
  {\bibfnamefont {B.}~\bibnamefont {Delley}}, \ and\ \bibinfo {author}
  {\bibfnamefont {T.}~\bibnamefont {Schmitt}},\ }\href
  {https://link.aps.org/doi/10.1103/PhysRevB.88.165103} {\bibfield  {journal}
  {\bibinfo  {journal} {Phys. Rev. B}\ }\textbf {\bibinfo {volume} {88}},\
  \bibinfo {pages} {165103} (\bibinfo {year} {2013})}\BibitemShut {NoStop}%
\bibitem [{\citenamefont {Herbrych}\ \emph {et~al.}(2018)\citenamefont
  {Herbrych}, \citenamefont {Kaushal}, \citenamefont {Nocera}, \citenamefont
  {Alvarez}, \citenamefont {Moreo},\ and\ \citenamefont
  {Dagotto}}]{Herbrych2018}%
  \BibitemOpen
  \bibfield  {author} {\bibinfo {author} {\bibfnamefont {J.}~\bibnamefont
  {Herbrych}}, \bibinfo {author} {\bibfnamefont {N.}~\bibnamefont {Kaushal}},
  \bibinfo {author} {\bibfnamefont {A.}~\bibnamefont {Nocera}}, \bibinfo
  {author} {\bibfnamefont {G.}~\bibnamefont {Alvarez}}, \bibinfo {author}
  {\bibfnamefont {A.}~\bibnamefont {Moreo}}, \ and\ \bibinfo {author}
  {\bibfnamefont {E.}~\bibnamefont {Dagotto}},\ }\href
  {https://doi.org/10.1038/s41467-018-06181-6} {\bibfield  {journal} {\bibinfo
  {journal} {Nat. Commun.}\ }\textbf {\bibinfo {volume} {9}},\ \bibinfo {pages}
  {3736} (\bibinfo {year} {2018})}\BibitemShut {NoStop}%
\bibitem [{\citenamefont {Patel}\ \emph {et~al.}(2019)\citenamefont {Patel},
  \citenamefont {Nocera}, \citenamefont {Alvarez}, \citenamefont {Moreo},
  \citenamefont {Johnston},\ and\ \citenamefont {Dagotto}}]{Patel2019}%
  \BibitemOpen
  \bibfield  {author} {\bibinfo {author} {\bibfnamefont {N.~D.}\ \bibnamefont
  {Patel}}, \bibinfo {author} {\bibfnamefont {A.}~\bibnamefont {Nocera}},
  \bibinfo {author} {\bibfnamefont {G.}~\bibnamefont {Alvarez}}, \bibinfo
  {author} {\bibfnamefont {A.}~\bibnamefont {Moreo}}, \bibinfo {author}
  {\bibfnamefont {S.}~\bibnamefont {Johnston}}, \ and\ \bibinfo {author}
  {\bibfnamefont {E.}~\bibnamefont {Dagotto}},\ }\href
  {https://doi.org/10.1038/s42005-019-0155-3} {\bibfield  {journal} {\bibinfo
  {journal} {Commun. Phys.}\ }\textbf {\bibinfo {volume} {2}},\ \bibinfo
  {pages} {64} (\bibinfo {year} {2019})}\BibitemShut {NoStop}%
\bibitem [{\citenamefont {Pizarro}\ and\ \citenamefont
  {Bascones}(2019)}]{Pizarro2019}%
  \BibitemOpen
  \bibfield  {author} {\bibinfo {author} {\bibfnamefont {J.~M.}\ \bibnamefont
  {Pizarro}}\ and\ \bibinfo {author} {\bibfnamefont {E.}~\bibnamefont
  {Bascones}},\ }\href
  {https://link.aps.org/doi/10.1103/PhysRevMaterials.3.014801} {\bibfield
  {journal} {\bibinfo  {journal} {Phys. Rev. Mater.}\ }\textbf {\bibinfo
  {volume} {3}},\ \bibinfo {pages} {014801} (\bibinfo {year}
  {2019})}\BibitemShut {NoStop}%
\bibitem [{\citenamefont {Suzuki}\ \emph {et~al.}(2015)\citenamefont {Suzuki},
  \citenamefont {Arita},\ and\ \citenamefont {Ikeda}}]{Suzuki2015}%
  \BibitemOpen
  \bibfield  {author} {\bibinfo {author} {\bibfnamefont {M.-T.}\ \bibnamefont
  {Suzuki}}, \bibinfo {author} {\bibfnamefont {R.}~\bibnamefont {Arita}}, \
  and\ \bibinfo {author} {\bibfnamefont {H.}~\bibnamefont {Ikeda}},\ }\href
  {https://link.aps.org/doi/10.1103/PhysRevB.92.085116} {\bibfield  {journal}
  {\bibinfo  {journal} {Phys. Rev. B}\ }\textbf {\bibinfo {volume} {92}},\
  \bibinfo {pages} {085116} (\bibinfo {year} {2015})}\BibitemShut {NoStop}%
\bibitem [{\citenamefont {Koley}(2018)}]{Koley2018}%
  \BibitemOpen
  \bibfield  {author} {\bibinfo {author} {\bibfnamefont {S.}~\bibnamefont
  {Koley}},\ }\href {\doibase 10.1088/2053-1591/aae94e} {\bibfield  {journal}
  {\bibinfo  {journal} {Mater. Res. Express}\ }\textbf {\bibinfo {volume}
  {6}},\ \bibinfo {pages} {016553} (\bibinfo {year} {2018})}\BibitemShut
  {NoStop}%
\bibitem [{\citenamefont {Slater}(1951)}]{Slater1951}%
  \BibitemOpen
  \bibfield  {author} {\bibinfo {author} {\bibfnamefont {J.~C.}\ \bibnamefont
  {Slater}},\ }\href {https://link.aps.org/doi/10.1103/PhysRev.82.538}
  {\bibfield  {journal} {\bibinfo  {journal} {Phys. Rev.}\ }\textbf {\bibinfo
  {volume} {82}},\ \bibinfo {pages} {538} (\bibinfo {year} {1951})}\BibitemShut
  {NoStop}%
\bibitem [{SM()}]{SM}%
  \BibitemOpen
  \href@noop {} {}\bibinfo {note} {For detailed information on sample
  preparation and characterization, see the Supplemental Material}\BibitemShut
  {NoStop}%
\bibitem [{\citenamefont {Dressel}\ and\ \citenamefont
  {Gr{\"u}ner}(2002)}]{Dressel2002}%
  \BibitemOpen
  \bibfield  {author} {\bibinfo {author} {\bibfnamefont {M.}~\bibnamefont
  {Dressel}}\ and\ \bibinfo {author} {\bibfnamefont {G.}~\bibnamefont
  {Gr{\"u}ner}},\ }\href {\doibase 10.1017/cbo9780511606168} {\emph {\bibinfo
  {title} {Electrodynamics of Solids: Optical Properties of Electrons in
  Matter}}}\ (\bibinfo  {publisher} {Cambridge University Press},\ \bibinfo
  {address} {Cambridge},\ \bibinfo {year} {2002})\BibitemShut {NoStop}%
\bibitem [{\citenamefont {Tanner}(2019)}]{Tanner2019}%
  \BibitemOpen
  \bibfield  {author} {\bibinfo {author} {\bibfnamefont {D.~B.}\ \bibnamefont
  {Tanner}},\ }\href {\doibase 10.1017/9781316672778} {\emph {\bibinfo {title}
  {Optical Effects in Solids}}}\ (\bibinfo  {publisher} {Cambridge University
  Press},\ \bibinfo {address} {Cambridge},\ \bibinfo {year} {2019})\BibitemShut
  {NoStop}%
\bibitem [{\citenamefont {Kuzmenko}(2005)}]{Kuzmenko2019}%
  \BibitemOpen
  \bibfield  {author} {\bibinfo {author} {\bibfnamefont {A.~B.}\ \bibnamefont
  {Kuzmenko}},\ }\href {\doibase 10.1063/1.1979470} {\bibfield  {journal}
  {\bibinfo  {journal} {Rev. Sci. Instrum.}\ }\textbf {\bibinfo {volume}
  {76}},\ \bibinfo {pages} {083108} (\bibinfo {year} {2005})}\BibitemShut
  {NoStop}%
\bibitem [{\citenamefont {Blaha}\ \emph {et~al.}(2018)\citenamefont {Blaha},
  \citenamefont {Schwarz}, \citenamefont {Madsen}, \citenamefont {Kvasnicka},
  \citenamefont {Luitz}, \citenamefont {Laskowski}, \citenamefont {Tran},\ and\
  \citenamefont {Marks}}]{Blaha}%
  \BibitemOpen
  \bibfield  {author} {\bibinfo {author} {\bibfnamefont {P.}~\bibnamefont
  {Blaha}}, \bibinfo {author} {\bibfnamefont {K.}~\bibnamefont {Schwarz}},
  \bibinfo {author} {\bibfnamefont {G.~K.~H.}\ \bibnamefont {Madsen}}, \bibinfo
  {author} {\bibfnamefont {D.}~\bibnamefont {Kvasnicka}}, \bibinfo {author}
  {\bibfnamefont {J.}~\bibnamefont {Luitz}}, \bibinfo {author} {\bibfnamefont
  {R.}~\bibnamefont {Laskowski}}, \bibinfo {author} {\bibfnamefont
  {F.}~\bibnamefont {Tran}}, \ and\ \bibinfo {author} {\bibfnamefont {L.~D.}\
  \bibnamefont {Marks}},\ }\href@noop {} {\emph {\bibinfo {title} {WIEN2k: An
  Augmented Plane Wave plus Local Orbitals Program for Calculating Crystal
  Properties}}}\ (\bibinfo  {publisher} {Vienna University of Technology},\
  \bibinfo {year} {2018})\BibitemShut {NoStop}%
\bibitem [{\citenamefont {Kim}\ \emph {et~al.}(2018)\citenamefont {Kim},
  \citenamefont {Kim}, \citenamefont {Sandilands}, \citenamefont {Sinn},
  \citenamefont {Lee}, \citenamefont {Son}, \citenamefont {Lee}, \citenamefont
  {Choi}, \citenamefont {Kim}, \citenamefont {Park}, \citenamefont {Jeon},
  \citenamefont {Kim}, \citenamefont {Park}, \citenamefont {Park},
  \citenamefont {Moon},\ and\ \citenamefont {Noh}}]{Kim2018}%
  \BibitemOpen
  \bibfield  {author} {\bibinfo {author} {\bibfnamefont {S.~Y.}\ \bibnamefont
  {Kim}}, \bibinfo {author} {\bibfnamefont {T.~Y.}\ \bibnamefont {Kim}},
  \bibinfo {author} {\bibfnamefont {L.~J.}\ \bibnamefont {Sandilands}},
  \bibinfo {author} {\bibfnamefont {S.}~\bibnamefont {Sinn}}, \bibinfo {author}
  {\bibfnamefont {M.-C.}\ \bibnamefont {Lee}}, \bibinfo {author} {\bibfnamefont
  {J.}~\bibnamefont {Son}}, \bibinfo {author} {\bibfnamefont {S.}~\bibnamefont
  {Lee}}, \bibinfo {author} {\bibfnamefont {K.-Y.}\ \bibnamefont {Choi}},
  \bibinfo {author} {\bibfnamefont {W.}~\bibnamefont {Kim}}, \bibinfo {author}
  {\bibfnamefont {B.-G.}\ \bibnamefont {Park}}, \bibinfo {author}
  {\bibfnamefont {C.}~\bibnamefont {Jeon}}, \bibinfo {author} {\bibfnamefont
  {H.-D.}\ \bibnamefont {Kim}}, \bibinfo {author} {\bibfnamefont {C.-H.}\
  \bibnamefont {Park}}, \bibinfo {author} {\bibfnamefont {J.-G.}\ \bibnamefont
  {Park}}, \bibinfo {author} {\bibfnamefont {S.~J.}\ \bibnamefont {Moon}}, \
  and\ \bibinfo {author} {\bibfnamefont {T.~W.}\ \bibnamefont {Noh}},\ }\href
  {https://link.aps.org/doi/10.1103/PhysRevLett.120.136402} {\bibfield
  {journal} {\bibinfo  {journal} {Phys. Rev. Lett.}\ }\textbf {\bibinfo
  {volume} {120}},\ \bibinfo {pages} {136402} (\bibinfo {year}
  {2018})}\BibitemShut {NoStop}%
\bibitem [{\citenamefont {Li}\ \emph {et~al.}(2016)\citenamefont {Li},
  \citenamefont {Kaushal}, \citenamefont {Wang}, \citenamefont {Tang},
  \citenamefont {Alvarez}, \citenamefont {Nocera}, \citenamefont {Maier},
  \citenamefont {Dagotto},\ and\ \citenamefont {Johnston}}]{Li2016}%
  \BibitemOpen
  \bibfield  {author} {\bibinfo {author} {\bibfnamefont {S.}~\bibnamefont
  {Li}}, \bibinfo {author} {\bibfnamefont {N.}~\bibnamefont {Kaushal}},
  \bibinfo {author} {\bibfnamefont {Y.}~\bibnamefont {Wang}}, \bibinfo {author}
  {\bibfnamefont {Y.}~\bibnamefont {Tang}}, \bibinfo {author} {\bibfnamefont
  {G.}~\bibnamefont {Alvarez}}, \bibinfo {author} {\bibfnamefont
  {A.}~\bibnamefont {Nocera}}, \bibinfo {author} {\bibfnamefont {T.~A.}\
  \bibnamefont {Maier}}, \bibinfo {author} {\bibfnamefont {E.}~\bibnamefont
  {Dagotto}}, \ and\ \bibinfo {author} {\bibfnamefont {S.}~\bibnamefont
  {Johnston}},\ }\href {https://link.aps.org/doi/10.1103/PhysRevB.94.235126}
  {\bibfield  {journal} {\bibinfo  {journal} {Phys. Rev. B}\ }\textbf {\bibinfo
  {volume} {94}},\ \bibinfo {pages} {235126} (\bibinfo {year}
  {2016})}\BibitemShut {NoStop}%
\bibitem [{\citenamefont {Lei}\ \emph {et~al.}(2011{\natexlab{b}})\citenamefont
  {Lei}, \citenamefont {Ryu}, \citenamefont {Frenkel},\ and\ \citenamefont
  {Petrovic}}]{Lei2011a}%
  \BibitemOpen
  \bibfield  {author} {\bibinfo {author} {\bibfnamefont {H.}~\bibnamefont
  {Lei}}, \bibinfo {author} {\bibfnamefont {H.}~\bibnamefont {Ryu}}, \bibinfo
  {author} {\bibfnamefont {A.~I.}\ \bibnamefont {Frenkel}}, \ and\ \bibinfo
  {author} {\bibfnamefont {C.}~\bibnamefont {Petrovic}},\ }\href {\doibase
  10.1103/PhysRevB.84.214511} {\bibfield  {journal} {\bibinfo  {journal} {Phys.
  Rev. B}\ }\textbf {\bibinfo {volume} {84}},\ \bibinfo {pages} {214511}
  (\bibinfo {year} {2011}{\natexlab{b}})}\BibitemShut {NoStop}%
\bibitem [{\citenamefont {Materne}\ \emph {et~al.}(2019)\citenamefont
  {Materne}, \citenamefont {Bi}, \citenamefont {Zhao}, \citenamefont {Hu},
  \citenamefont {Amigó}, \citenamefont {Seiro}, \citenamefont {Aswartham},
  \citenamefont {Büchner},\ and\ \citenamefont {Alp}}]{Materne2019}%
  \BibitemOpen
  \bibfield  {author} {\bibinfo {author} {\bibfnamefont {P.}~\bibnamefont
  {Materne}}, \bibinfo {author} {\bibfnamefont {W.}~\bibnamefont {Bi}},
  \bibinfo {author} {\bibfnamefont {J.}~\bibnamefont {Zhao}}, \bibinfo {author}
  {\bibfnamefont {M.~Y.}\ \bibnamefont {Hu}}, \bibinfo {author} {\bibfnamefont
  {M.~L.}\ \bibnamefont {Amigó}}, \bibinfo {author} {\bibfnamefont
  {S.}~\bibnamefont {Seiro}}, \bibinfo {author} {\bibfnamefont
  {S.}~\bibnamefont {Aswartham}}, \bibinfo {author} {\bibfnamefont
  {B.}~\bibnamefont {Büchner}}, \ and\ \bibinfo {author} {\bibfnamefont
  {E.~E.}\ \bibnamefont {Alp}},\ }\href
  {https://link.aps.org/doi/10.1103/PhysRevB.99.020505} {\bibfield  {journal}
  {\bibinfo  {journal} {Phys. Rev. B}\ }\textbf {\bibinfo {volume} {99}},\
  \bibinfo {pages} {020505} (\bibinfo {year} {2019})}\BibitemShut {NoStop}%
\end{thebibliography}%

\end{document}